\documentclass[prb,nofootinbib,amsfonts,tightenlines,twocolumn]{revtex4-2}
\usepackage{graphicx}
\usepackage{dcolumn}
\usepackage{bm}
\usepackage{caption}
\usepackage{subcaption}
\usepackage{mathrsfs}
\usepackage{dsfont}
\usepackage{physics}
\usepackage{natbib}
\usepackage{xcolor}
\usepackage{wrapfig}\usepackage{upgreek}
\usepackage[utf8]{inputenc}
\usepackage{xpatch}
\usepackage{amsmath}

\usepackage{soul}
\usepackage[normalem]{ulem}

\begin{document}

\title{Thermoelectric response of Josephson junction: from ballistic to disordered}
\author{Aabir Mukhopadhyay} 
\email{aabir.riku@gmail.com}
\author{Sourin Das}
\email{sourin@iiserkol.ac.in, \\ ORCID ID : 0000-0002-8511-5709}
\affiliation{Indian Institute of Science Education \& Research Kolkata,
Mohanpur, Nadia - 741 246, 
West Bengal, India}
\date{\today}

\begin{abstract}
It is known that Josephson junction (JJ) hosting scattering centers with energy dependent scattering amplitudes which breaks the $\omega\rightarrow-\omega$ symmetry (where $\omega$ is the excitation energy of electron about the Fermi level) exhibits finite thermoelectric response. In contrast, here we show that even in a ballistic JJ this symmetry is broken and it leads to a non-zero thermoelectric response when the junction length is of the order of coherence length of the superconductor and the corresponding thermoelectric coefficient confirms to the universal sinusoidal dependence on $\phi_{12}$, where $\phi_{12}$ is the superconducting phase bias. In presence of multiple scatterers in the junction region, we have numerically shown that the sign of the even-in-$\phi_{12}$  part of the thermoelectric coefficient fluctuates violently from one disorder configuration to another hence averaging to vanishingly small values while the odd part tends towards the universal sinusoidal dependence on $\phi_{12}$ as we approach the large disorder limit under disorder averaging.
\end{abstract}

\maketitle
\section{Introduction}
Thermoelectric response of a hybrid junction between two normal metals in the mesoscopic regime has been discussed extensively both theoretically and experimentally \cite{sivan_PRB_33_551,staring_EPL_22_57, moller_PRL_81_5197, scheibner_PRL_95_176602,ludoph_PRB_59_12290, reddy_AAAS_315_1568, widawsky_NL_12_354, thierschmann_NN_10_854, shankouri_ARMR_41_399, sanchez_PRB_83_085428, beenakker_PRB_46_9667, entin_PRB_82_115314, sanchez_PRB_84_201307, jordan_PRB_87_075312, brandner_PRL_110_070603}. Whereas, analogous situation comprising of a junction of superconductors is a less explored topic though discussion of thermoelectric response of superconductor has a long history. Such set-ups are of great importance because of the possibility of its applications in improving the efficiency of thermoelectric generator by strongly suppressing Ohmic losses \cite{kolenda_PRL_116_097001, kolenda_PRB_95_224505, shimizu_NC_10_825, tan_NC_12_138, fornieri_NN_12_944, giazotto_RMP_78_217}. 

In 1944, Ginzberg \cite{ginzburg_JPUSSR_8_148, ginzburg_RMP_76_981} showed that a temperature gradient in a bulk superconductor leads to a finite normal current response, though this current gets completely cancelled by a counter flow of supercurrent in a homogeneous isotropic superconductor which make it impossible to detect the thermoelectric response in isolation. This fact lead him to theoretically explore the possibilities of anisotropic and inhomogeneous superconductor for the detection of the thermoelectric effect. Since then, various theoretical study\cite{ginzburg_SST_4_S1, ginzburg_PCS_235_3129, marinescu_PRB_55_11637, galperin_ZETF_66_1387, galperin_PRB__65_064531, virtanen_APA_89_625} has been conducted exploring possibilities of detection of thermoelectric response of superconductors in anisotropic and inhomogeneous situations. Experimental study in this direction goes back all the way to 1920's \cite{Falco1981book, borelius_PKNAW_34_1365, burton_Nature_136_141, keesom_Physica_5_437, casimir_Physica_13_33, pullan_PRSLSA_217_280, harlingen_PRB_21_1842, kartsovnik_JETPL_33_7, fornieri_NN_11_258} and this topic has been revisited in the recent past in an interesting work by Shelly et.al.\cite{connor_SA_2_e1501250}. The discovery of Josephson effect \cite{josephson_PL_1_251} in 1962 provided a natural setting for exploring  thermoelectric response for a inhomogeneous superconductor. Later in 1997, Guttman and Bergman made an attempt to theoretically explore the thermoelectric response of a JJ in a tunnel Hamiltonian approach  \cite{guttman_PRB_55_12691}.

Pershoguba and Glazman \cite{pershoguba_PRB_99_134514} have carried out an elaborate study on the possibility of generating thermometric current across a junction between two quasi-one dimensional superconductors, which goes beyond the tunneling limit and also discussed the relevance of the odd and the even part of the Josephson current as a function of the superconducting phase bias $\phi_{12}$ owing to scattering in junction region which breaks the $\omega \rightarrow -\omega$ symmetry. In this regard, the helical edge state of two dimensional topological insulators pose an interesting and cleane testing ground for such theoretical study which hosts one-dimensional Josephson junction\cite{hart_NP_10_638}. Thermal response of quantum hall edge has already being studies  in experiment\cite{banerjee_Nat_545_75} and hence an similar experimental set-up involving the spin Hall edge may not be far in the future. Recent theoretical studies have explored the possibility of inducing thermoelectric effect in helical edge state-based Josephson junction involving either an anisotropic ferromagnetic barrier\cite{gresta_PRL_123_186801, marchegiani_APL_117_212601} or a three-terminal geometry\cite{blasi_PRL_124_227701, blasi_PRB_103_235434, blasi_PRB_102_241302}. In this work we show that thermoelectric effect can exist in the HES of QSH even in a simplest case of two terminal ballistic JJ owing to breaking of the $\omega \rightarrow -\omega$ symmetry of the quasi-particle transmission probabilities across the junction at finite length. We argue that this is generic to ballistic JJ and is not specific to HES. Lastly, it is worth noting that the use of thermal transport for probing quantum states has been much in pursuit in contemporary science\cite{li_MRSB_45_348} and hence such a discussion is quite timely.

The paper is organized as follows. In Section \ref{system_HES} we described the JJ based on HES of a 2D QSH state and in Section \ref{left_righth_symmetry} we discussed how a long ballistic JJ can break the $\omega \rightarrow -\omega$ symmetry and hence resulting in thermoelectric response which also survives in presence of disorder. In Section \ref{even_and_odd_section} we extend our discussion to the odd-in-$\phi_{12}$ and even-in-$\phi_{12}$ part of the thermoelectric conductance and shown that minimal breaking of the $\omega \rightarrow -\omega$ symmetry is not enough to induce an even-in-$\phi_{12}$ contribution.We have also argued that the presence of thermoelectric response through the breaking of $\omega \rightarrow -\omega$ symmetry is not unique to HES, rather it is a generic property of a JJ.



\section{Ballistic Josephson junction in helical edge state}
\label{system_HES}

\begin{figure}[]
	\includegraphics[width=0.4\textwidth]{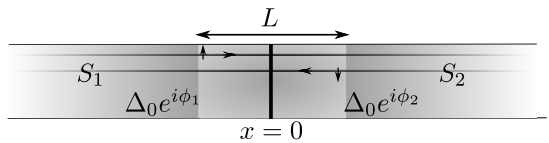}
	\caption{Schematic of the Josephson junction set-up in a Helical edge state.}
	\label{set_up}
\end{figure}

\begin{figure*}[t!]
	\includegraphics[width=\textwidth]{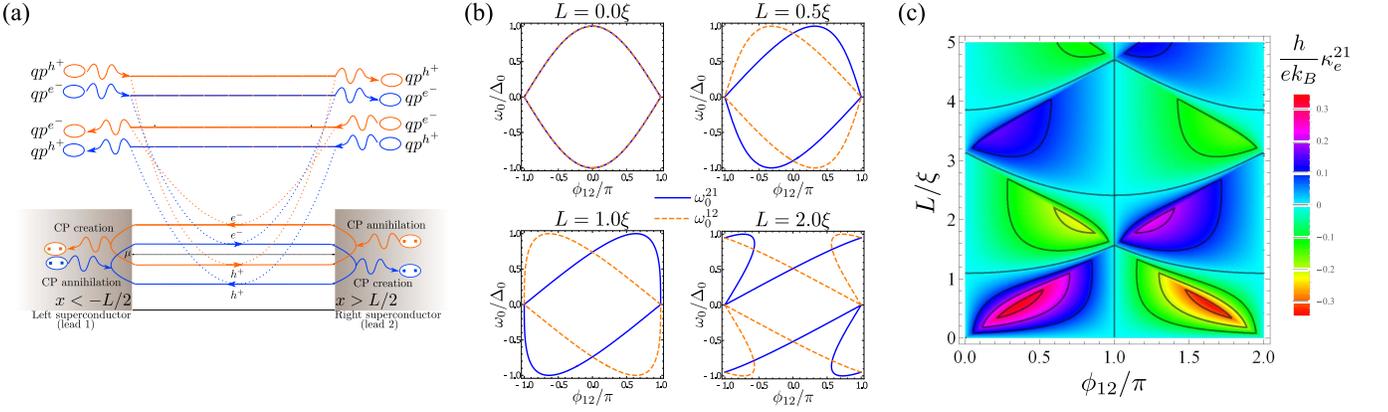}
	\caption{(a) Pictorial representation of below the gap $(\omega<\Delta_0)$ tunneling of Cooper pairs (CP) form left(right) to right(left) via two different Andreev bound states $\omega_0^{21}$ (indicated by blue lines) and $\omega_0^{12}$ (indicated by orange lines). The dotted lines represent the fact that tunnelling of the quasielectrons (quasiholes) above the gap $(\omega>\Delta_0)$ across the junction are in correspondence with distinct bound states, as can be noted from the poles of the quasielectron (quasihole) transmission probabilities $\mathcal{T}_{ee}^{21}$ and $\mathcal{T}_{ee}^{12}$ (or $\mathcal{T}_{hh}^{12}$ and $\mathcal{T}_{hh}^{21}$). (b) Two types of Andreev bound states as a function of superconducting phase difference $\phi_{12}$, are plotted for different values of junction lengths where $\xi=\hbar v_F/\Delta_0$ is the superconducting coherence length. 
	(c) Density plot for the thermoelectric coefficient $\kappa^{21}$ of a ballistic JJ based on the edge states of a quantum spin Hall insulator in proximity to a s-wave superconductor, as a function of superconducting phase bias  $\phi_{12}$ and junction length $L$. The average temperature of the junction is considered to be $k_BT=0.5 \Delta_0$.}
	\label{ABSeps}
\end{figure*}

\begin{figure*}[t]
	\centering
	\includegraphics[width=0.9 \textwidth]{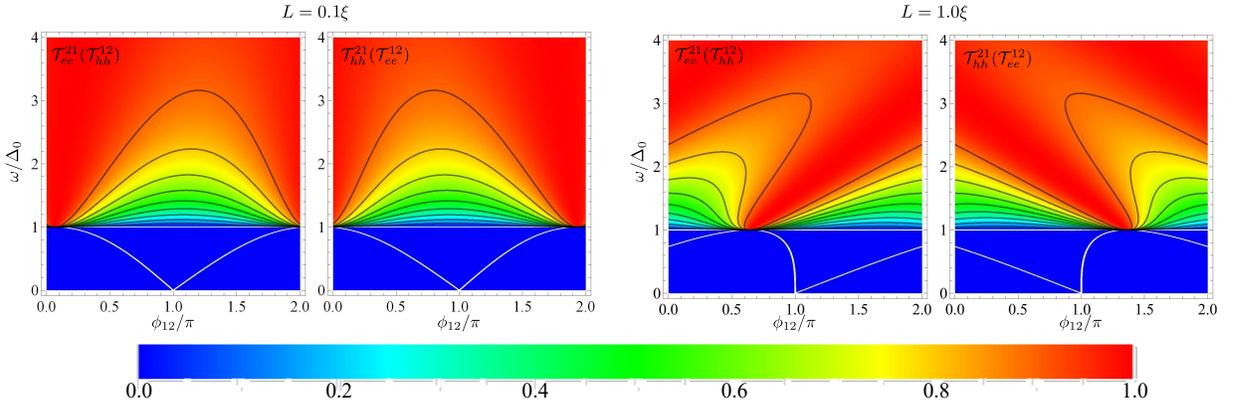}
	\caption{Different transmission probabilities of the quasiparticles through a ballistic Josephson junction based on the helical edge state of a quantum spin Hall insulator, in the space of energy ($\omega$) and superconducting phase-difference ($\phi_{12}$) for different values of junction length ($L$). A clear asymmetry between the electron and hole transmission probability from left (right) to right(left) develops as we increase the length of the junction. The plot in energy window $(\vert\omega\vert<\Delta_0)$ signifies the evolution of the pole (location of ABS) of the transmission amplitude as a function of $\phi_{12}$.}
	\label{transmission}
\end{figure*}

We first consider a JJ based on 1D Dirac fermions in proximity to a s-wave superconductor, realized in a HES of QSH insulator\cite{fu_PRB_79_161408, fu_PRL_100_096407, calzona_arxiv_1909_06280} because of its algebraic simplicity. Later we will also explore the case of quadratic dispersion. The junction is considered to be of length  $L$ laying over the region $|x|<L/2$. The proximitized region of the edge are described by the Bogoliubov-de Gennes (BdG) Hamiltonian in the Nambu basis\cite{fu_PRB_79_161408, fu_PRL_100_096407} $  ([(\psi_{\uparrow},\psi_{\downarrow}),(\psi_{\downarrow}^{\dagger},-\psi_{\uparrow}^{\dagger})]$) as
\begin{equation}
	\mathcal{H}=(-i\hbar v_F \partial_x \sigma_z-\mu)\tau_z +\Delta(x) (\cos \phi_r \tau_x -\sin \phi_r \tau_y);	\label{DiracHamiltonian}
\end{equation}
where $\sigma$ and $\tau$ are the Pauli matrices representing spin and particle-hole degrees of freedom respectively. The superconducting pairing potential is given by $\Delta(x)=\Delta_0 [\Theta(-x-L/2)+\theta(x-L/2)]$ such that it defines a superconductor-normal-superconductor junction (SNS). Superconducting leads ($S_{r}$) are identified as $r \in \{1,2\}$ (left lead being $r=1$ and the right being $r=2$) and $\phi_r$ are the corresponding superconducting phases (see Fig. \ref{set_up}); $\mu$ is the chemical potential throughout the edge and $v_F$ is the corresponding Fermi velocity. We also consider the doping to be finite $(\mu\neq 0)$ and  the length of the junction $L$ to be comparable to the superconducting coherence length $\xi=\hbar v_F/\Delta_0$ such that $(p_e-p_h)L/\hbar$ can be of the order of unity for energies $\leq \Delta_0 $, where $p_{e/h}=\hbar k_{e/h}$ are the quasi-particle and the corresponding quasi-hole momenta for particles in the junction region. Note that, in general, for highly doped superconductor with quadratic dispersion $p_e \approx p_h$ and hence such phases are generally neglected. On the other hand, if such phase accumulation becomes of the order of unity, it naturally leads to breaking of $\omega \rightarrow -\omega$ symmetry of the excited Bogoluibov quasielectron and quasihole transmission probabilities individually across the junction leading to finite thermoelectric effect, though the sum of the two does respect the symmetry.

\section{$\omega \rightarrow -\omega$ symmetry breaking and Andreev bound state of ballistic Josephson junction}
\label{left_righth_symmetry}
Let us consider a right moving electron-like quasiparticle which starts its journey at $x=-L/2$ and propagates through the normal region and reaches at $x=L/2$. It Andreev reflects back as a hole with an uni-modular amplitude by creating a Cooper pair in the superconducting lead 2 (S2). The reflected hole then travels through the normal region and reaches back to $x=-L/2$. It then suffers a second Andreev reflection hence annihilating a Cooper pair at superconducting lead 1 (S1) and completing a closed loop journey resulting in shuttling of a single Copper pair from S1 to S2 [See Fig.\ref{ABSeps} (a)]. This process involving a right-moving electron and a left-moving hole can be directly related to formation of a ABS at the JJ where  the ABS energy is given by $\omega_0^{21} =\pm \Delta_0 \left\lvert\cos (\frac{k_e(\omega_0^{21})-k_h(\omega_0^{21})}{2}L-\frac{\phi_{12}}{2})\right\rvert$ where $\phi_{12}=\phi_2-\phi_1$ and $k_{e,h}(\omega_0)=(\mu \pm \omega_0)/(\hbar v_F)$(See Appendix \ref{Appendix_clean_junction}). Similarly, if a right-moving hole-like quasiparticle starts from $x=-L/2$ and completes the cycle after two Andreev reflections, it will transfer a Cooper pair from S2 to S1 [See Fig.\ref{ABSeps}(a)] and the corresponding ABS will be formed at energies $\omega_0^{12} =\pm \Delta_0 \left\lvert\cos (\frac{k_e(\omega_0^{12})-k_h(\omega_0^{12})}{2}L+\frac{\phi_{12}}{2})\right\rvert$ (See Appendix \ref{Appendix_clean_junction}). Note that, the $\omega_0^{21}$ and $\omega_0^{12}$ transform into one another as 
$\phi_{12} \rightarrow -\phi_{12}$. The ABS energies $\omega_0^{21}$ and $\omega_0^{12}$ are shown as a function of $\phi_{12}$ for different values of the junction length $L$ in Fig.\ref{ABSeps} (b). The important point to note here is the fact that,  for finite L, the degeneracy between $\omega_0^{21}$ and $\omega_0^{12}$ is lifted whenever $\phi_{12}\neq 0,\pi$ and this fact leads to an asymmetry between the transmission probability of electron- (hole-) like  BdG quasiparticle above the gap, incident on the junction from the left and right hence leading to finite thermoelectric response.  

Now, for analyzing the implication of degeneracy lifting of ABS on  the thermoelectric effect of the junction, we start by calculating the scattering amplitude for Bogoliubov quasiparticle above the gap ($\omega>\Delta_0$ ) across the JJ. It is straightforward to match the plane wave solutions of the BdG equation to obtain the transmission probabilities across the JJ (from $S1$ to $S2$) as described by Eq. \ref{DiracHamiltonian} are given by (see Appendix \ref{Appendix_clean_junction})
\begin{align}
	\mathcal{T}_{ee}^{21} &= \mathcal{T}_{hh}^{12}  &=\dfrac{\omega^2-\Delta_0^2}{\omega^2-\Delta_0^2 \cos^2 \left( \frac{k_e-k_h}{2}L-\frac{\phi_{12}}{2} \right)},	\label{T_ee^21}\\
	\mathcal{T}_{hh}^{21} &= \mathcal{T}_{ee}^{12}  &=\dfrac{\omega^2-\Delta_0^2}{\omega^2-\Delta_0^2 \cos^2 \left( \frac{k_e-k_h}{2}L+\frac{\phi_{12}}{2} \right)},   \label{T_hh^21}
\end{align}
while $\mathcal{T}_{he}^{21}=\mathcal{T}_{eh}^{21}=\mathcal{T}_{he}^{12}=\mathcal{T}_{eh}^{12}=0$. Quasiparticle transmission probabilities through a ballistic JJ is shown in FIG. \ref{transmission} for two different lengths of the junction.  Here $\mathcal{T}_{q'q}^{ji}$ denote the transmission probability of an $q$-like QP ($q=e,h$) from lead $Si$ to a  $q'$-like QP in lead $Sj$. Note that, the tunneling of an electron- (hole-) like QP from S1 to S2 (S2 to S1) is in correspondence with the ABS having energy $\omega_0^{21}$ while the tunneling of a hole- (electron-) like QP from S1 to S2 (S2 to S1) is in correspondence with the ABS having energy $\omega_0^{12}$ [See Fig. \ref{ABSeps}(a)] which is apparent from the fact that the poles of the transmission amplitudes for these two processes coincides with the corresponding ABS energies. Within linear response theory, thermoelectric coefficient of a JJ can be defined in terms of the transmission probabilities as\cite{pershoguba_PRB_99_134514}
\begin{align}
	\kappa^{21} = \left[\dfrac{e}{h}\int_{\Delta_0}^{\infty} d\omega \dfrac{\omega}{\sqrt{\omega^2-\Delta^2}} [i_{e}^{21}-i^{21}_{h}] \dfrac{d\mathrm{f}(\omega,T)}{dT}\right]_{T=T_{\text{avg}}}	\label{thermoelectricCoefficient}
\end{align}
where $i^{21}_{e}=(\mathcal{T}_{ee}^{21}-\mathcal{T}_{he}^{21})$, $i^{21}_{h}=(\mathcal{T}_{hh}^{21}-\mathcal{T}_{eh}^{21})$, $e$ is the electronic charge, $\mathrm{f (\omega,T)}$ is the Fermi distribution function at temperature $T$ and $T_{\text{avg}}$ is the average temperature of the junction. Note that, in the limit $L \rightarrow 0$, $\kappa^{21}$ is zero.

The integration in Eq. (\ref{thermoelectricCoefficient}) can be done numerically and $\kappa^{21}$ can be obtained as a function of superconducting phase difference $\phi_{12}$ and junction length $L$ which is plotted as a density plot in FIG. \ref{ABSeps}(c). In case of HES, owing to its linear dispersion, the value of the overall chemical potential $\mu$ does not effect the calculations for the ballistic case.

To obtain an estimate of the extremum values of thermoelectric conductance for a single channel ballistic junction, we perform a numerical scan over the parameter space of $\phi_{12}$ and $L$ for a given temperature of $k_BT_{\text{avg}}=0.5\Delta_0$. We found that the maximum of (minimum of) $|\kappa^{21}_e| \approx\,0.3438 \,ek_B/h$ $(\approx 1.477 nA/K)$ is obtained at a junction length $L \approx 0.555 \xi$ for $\phi_{12}\approx 0.353 \pi$ (minimum at $\phi_{12} \approx 1.647 \pi$).

\begin{figure*}[t]
	\includegraphics[width=1.0 \textwidth]{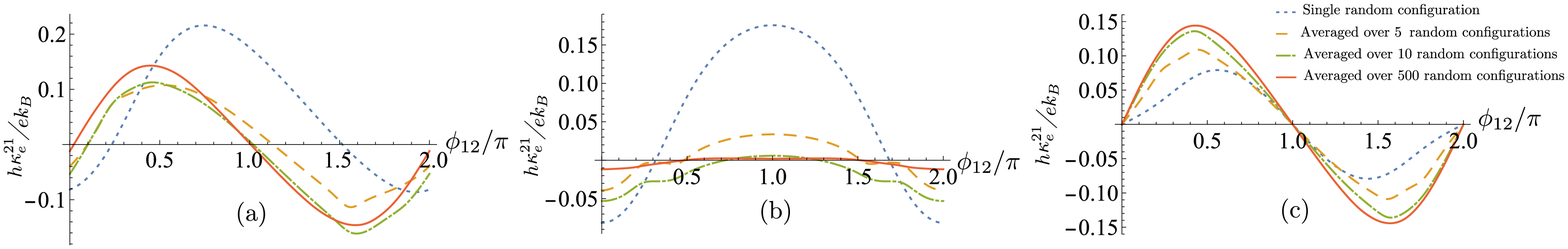}
	\caption{Thermoelectric conductance of a Josephson junction based on helical edge state of quantum spin Hall insulator in presence of four random scattering centers. (a) Total thermoelectric conductance (b) the part of thermal conductance that is even in $\phi_{12}$ (c) the part of conductance that is odd in $\phi_{12}$, for single random configuration of scattering centers and after averaging over different numbers of random configurations.}
	\label{even_odd}
\end{figure*}

As we can see from FIG. \ref{transmission}, for a ballistic JJ, in general for any given value of $\phi_{12}$ and at an energy $\omega>\Delta_0$ the quasiparticle transmission probabilities $\mathcal{T}_{ee}^{21}$ and $\mathcal{T}_{hh}^{21}$ are different if the length of the junction $L$ is comparable to the superconducting coherence length (i.e. when we are not in the short junction limit). Note that, the difference between these quantities at a given $\omega$ is maximum in the neighborhood of $\omega=\Delta_0$ and it decreases as we go higher in $\omega$, although non-monotonically. Additionally, one must notice, the thermoelectric effect identically vanishes both at $\phi_{12}=0$ and $\pi$ which are the time reversal symmetric points (See FIG. \ref{ABSeps}(c)).

\section{Even-in-$\phi_{12}$ and Odd-in-$\phi_{12}$ part of the thermoelectric response and the effect of disorder}
\label{even_and_odd_section}
Presence of scatter within the junction region, which breaks the $\omega \rightarrow -\omega$ symmetry, not only leads to a finite thermoelectric conductance, but also results in deviation from thermoelectric conductance being odd in $\phi_{12}$\cite{pershoguba_PRB_99_134514}. As discussed above, a JJ of finite length also breaks the $\omega \rightarrow -\omega$ symmetry hence it is curious if this minimal symmetry breaking can result in such a deviation, i.e. the thermoelectric response can be written as a liner sum of an even-in-$\phi_{12}$ part and an odd-in-$\phi_{12}$ part.

It is straightforward to check that the expression for thermoelectric conductance in the ballistic limit, obtained from Eq. \ref{T_ee^21}, \ref{T_hh^21} and \ref{thermoelectricCoefficient} is an odd function of $\phi_{12}$, independent of the length of the junction. This implies that the breaking of $\omega \rightarrow -\omega$ symmetry via $k_e\neq k_h$ (as discussed in the previous section) does not lead to any contribution to the thermoelectric response which is even in $\phi_{12}$. Further, we calculate the thermoelectric conductance in presence of a single localized scatterer which is positioned at an arbitrary point within the junction region and we assume that the scattering matrix corresponding to the scatterer has no energy dependence. The expression for the thermoelectric conductance in this case is given below,
\begin{widetext}
\begin{equation}
    \kappa^{21} = \left[\dfrac{e}{h}\int_{\Delta_0}^{\infty} d\omega \dfrac{\omega}{\sqrt{\omega^2-\Delta^2}} \left[ \dfrac{4 \tau \left((1-\tau) \sin{((k_e-k_h)L(m-n))}+ \sin{((k_e-k_h)L)} \right)\sin{\phi_{12}} \sinh2{\theta}}{\Omega \Omega^*} \right] \dfrac{d\mathrm{f}(\omega,T)}{dT}\right]_{T=T_{\text{avg}}},
    \label{m_n_junction}
\end{equation}
\end{widetext}
where, $\Omega=(1-\tau) \cos{\left((k_e-k_h)L(m-n)\right)} + \cos{\left( (k_e-k_h)L-2i\theta \right)}-\tau \cos{\phi_{12}}$, $\theta=\text{arccosh}{\omega/\Delta_0}$, $\tau$ is the normal state transmission probability across the scatterer and the position of the scattering center divides the junction region in the ratio $m:n$ ($m,n\leq 1$ and $m+n=1$). All other notations have their usual meanings as discussed before. Eq. \ref{m_n_junction} clearly shows that the thermoelectric response in this case also, is odd in $\phi_{12}$. Hence, our study establishes the fact that the minimal breaking of $\omega \rightarrow - \omega$ symmetry for a finite length ballistic junction (or in presence of a single scatterer which does not break the $\omega \rightarrow -\omega$ symmetry) is sufficient to induce thermoelectric response across the JJ, though it is not enough to induce an even-in-$\phi_{12}$ contribution to the thermoelectric conductance.

\begin{figure*}[t]
	\includegraphics[width=1.0 \textwidth]{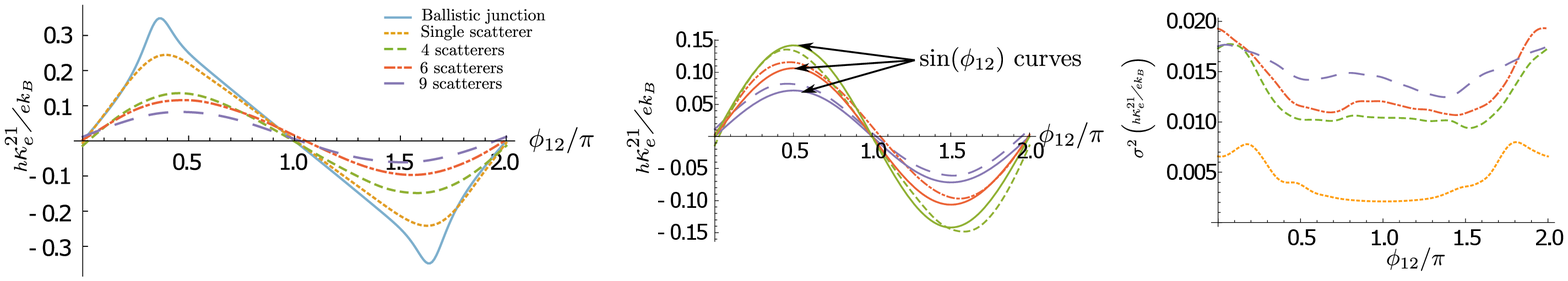}
	\caption{Disordered averaged mean value (left figure) and the variance (right figure) of the thermoelectric coefficient $\kappa^{21}$ of a S-TI-S junction based on the edge states of a quantum spin Hall insulator with proximity to a s-wave superconductor, are plotted as a function of superconducting phase difference $\phi_{12}$ . The average temperature of the junction is considered to be $k_BT=0.5 \Delta_0$ and the overall chemical potential to be $\mu=10\Delta_0$. Length of the junction is considered to be $L=0.555 \xi$ where $\xi$ is the superconducting coherence length. Average is done over 500 disorder configurations. The middle plot show that, as we increase the number of scatterers, the curves for thermoelectric conductance tend to a $\sin{(\phi_{12})}$ curves (solid lines) with an amplitude (Max($\kappa_e^{12}$)-Min($\kappa_e^{21}$))/2.}
	\label{multipleDisorders}
\end{figure*}

Now, if we consider a situation comprising of more than one such scatterer, then the effective scattering matrix describing the collection of scatterers will become energy dependent and in general will also break the $\omega \rightarrow -\omega$ symmetry, resulting in an even-in-$\phi_{12}$ contribution to the thermoelectric conductance as expected \cite{pershoguba_PRB_99_134514}. The even-in-$\phi_{12}$ part of the thermoelectric conductance is proportional to $(\tau_{\omega} -\tau_{-\omega})$, where $\tau_{\omega}$ is the normal state transmission probability across the junction at an energy $\omega$, and thus can vary drastically (both in amplitude and in sign) for different disorder configurations for a given $\phi_{12}$. Hence, averaging over random configurations results in vanishingly small values of the even part. Next we perform a numerical calculation to analyze the effect of averaging over a large number of disorder configurations in presence of multiple scatterers. To begin with, we consider four scattering centers represented by four energy-independent scattering matrices placed at random positions inside the junction region. Transmission probabilities of the scattering matrices are chosen randomly from a one-sided Gaussian distribution with a mean of $95\%$ and standard deviation of $5\%$. All the phase freedom of the disorders have been chosen randomly from a Gaussian distribution with a mean of $0$ and standard deviation $0.05 \pi$. We have fixed the length of the junction to be $L=0.555 \xi$, the value at which we get maximum thermoelectric conductance (which occurs for $\phi=0.353 \pi$) for a ballistic JJ. It can be seen clearly from FIG. \ref{even_odd} that averaging over as-small-as 10 configurations already shows a convergence towards an odd-in-$\phi_{12}$ behaviour while the even-in-$\phi_{12}$ part is strongly suppressed.  It is interesting to note that, in absence of an averaging (corresponding to a fixed quenched disorder configuration), for certain range of values of $\phi_{12}$, the even part can be the dominant contribution in the net thermal conductance (See FIG. \ref{even_odd}).

Now we extend the numerical analysis to a larger number of scattering centers. The scattering centers are modeled as before and the length of the junction is fixed at $L=0.555 \xi$. For a given number of scattering centers, disorder average is done over 500 configurations where we have checked that beyond this, there is negligible variation of the result. The mean and the variance of the thermoelectric conductances are plotted as a function of the superconducting phase difference $\phi_{12}$ in FIG.\ref{multipleDisorders}. Note that, in presence of a single scatterer, the variance of the thermoelectric conductance is smallest because in this case there is no even-in-$\phi_{12}$ part of the thermoelectric conductance. We have also observed that, in general, the variance is relatively lower in the neighborhood of $\phi_{12}=\pi$ rather than in the neighborhood of $\phi_{12}=0$ or $2\pi$. To conclude, the plot for thermoelectric conductance after averaging tend to reduce to the universal sinusoidal dependence of $\phi_{12}$ as the number of scatterers within the junction region increases (see the middle figure of FIG. \ref{multipleDisorders}). This is due to the fact that, with increasing opacity of the JJ, the $\phi_{12}$ sensitivity of the thermoelectric conductance via the poles of the quasiparticle transmission probabilities decreases and the major contribution comes from the explicit $\sin{(\phi_{12})}$ factor in the numerator.


\begin{figure*}[t]
	\includegraphics[width=0.7 \textwidth]{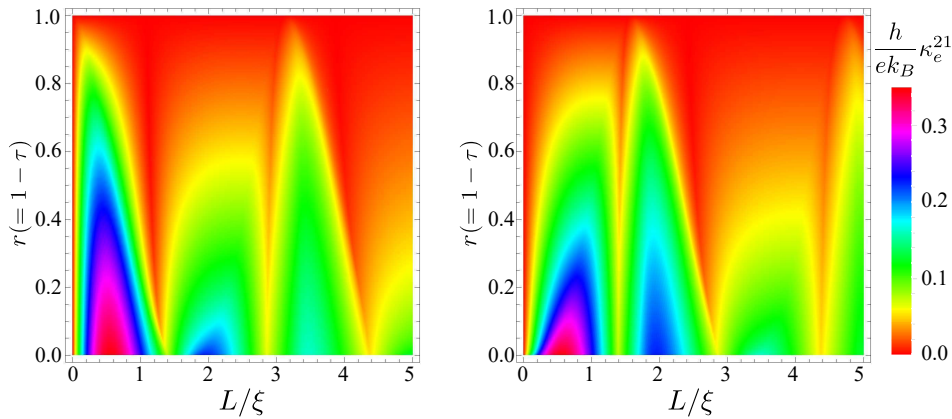}
	\caption{The maximum possible thermoelectric coefficient of a JJ with (left) s-wave and (right) p-wave superconductivity for a given junction length and normal state reflection probability $r$ (as calculated wit the analytic approximation $\mu>>\Delta_0,k_BT$). Note that the scatterer is assumed to be energy independent and is placed at the middle of the junction. The parameters are assumed to be $\mu=100\Delta_0$ and $k_BT=0.5\Delta_0$.}
	\label{max_thermo}
\end{figure*}

\section{Discussion}
Occurrence of thermoelectric effect through the breaking of $\omega \rightarrow -\omega$ symmetry for a ballistic long JJ is not specific to the HES. 1D JJ with quadratic dispersion and with s-wave or p-wave superconductivity should also demonstrate such a response. Of course, in the high doping limit, the thermoelectric coefficient should reduce to the results obtained in the paper when linearized about the Fermi energy. Thus, the thermoelectric response is a generic property of any ballistic JJ with junction length of the order of the superconducting coherence length. However, with the increasing opacity of the JJ for junction length less than the superconducting coherence length, the ABS energies tend to move towards the zero energy for p-wave superconductivity due to the presence of Majorana fermions. Whereas, for a JJ with s-wave superconductivity within the same limit, the ABS energies tend to move towards the continuum with increasing opacity of the junction. This fact manifests itself in the thermoelectric conductance via the poles of the quasiparticle transmission probabilities. Also, for JJ with junction length longer than the superconducting coherence length, the states from the continuum spectrum tend to leak into the superconducting gap, thereby changing the details of the thermoelectric coefficient.

Further, to check if the s-wave or the p-wave leads to a larger thermoelectric coefficient for a given junction length we perform an analysis where we have placed an energy-independent scatterer at the middle of a JJ, and plotted the maximum possible thermoelectric conductance (scanned over all values of $\phi_{12}$) for a given junction length $L$ and given transparency of the scatterer (normal state transmission probability $\tau$) as shown in FIG. \ref{max_thermo}. We have performed this study within the approximation $\mu>>\Delta_0, k_BT$ (See Appendix \ref{Appendix_middle_scatterer}). From these results we can conclude that in general, there is no distinguishable pattern in the thermoelectric coefficient for the case of s-wave and p-wave superconductivity.

As far as the possible strategy for the measurement of the thermoelectric current is concerned, it cannot be measured in isolation as it will always be accompanied by the finite temperature Josephson current. However, there may be ways to measure the thermoelectric coefficient indirectly. For example, consider a situation where a JJ is initially maintained at an equilibrium temperature $T$. The current that is obtained, is totally the Josephson current $\mathtt{S}_{(T,T)}=I_J$, where the first (second) subscript corresponds to the temperature of the left (right) lead S1 (S2). Now, if $S1$ is raised to temperature $T+\Delta T$, then the corresponding total current will be a sum of the Josephson current and the thermoelectric current $\mathtt{S}_{(T+\Delta T, T)}=I_J-\Delta I_J + \kappa^{21}_e \Delta T$, where $\Delta I_J$ is the variaation in the Josephson current due temperature bias. Next, consider the situation where $S1$ is kept at temperature $T$ while $S2$ is raised to temperature $T+\Delta T$, then the corresponding total current will be $\mathtt{S}_{(T, T+\Delta T)}=I_J-\Delta I_J - \kappa^{21}_e \Delta T$. Now if, $(2\mathtt{S}_{(T,T)}-(\mathtt{S}_{(T+\Delta T, T)}+\mathtt{S}_{(T, T+\Delta T)}))/2\mathtt{S}_{(T,T)}<< 1$ then a measurement of the ratio, $(\mathtt{S}_{(T+\Delta T, T)}-\mathtt{S}_{(T, T+\Delta T)})/2\Delta T$ will provide the thermoelectric coefficient. Note that, a similar strategy involving $\phi_{12}\rightarrow -\phi_{12}$ rather than involving $\Delta T \rightarrow -\Delta T$ is difficult to implement due to the presence of even-in-$\phi_{12}$ part of the thermoelectric coefficient.

\textit{\underline{Acknowledgment}}: A.M. thanks Vivekananda Adak for helpful discussions. We thank Chris Olund, Sergey Pershoguba and Erhai Zhao for useful communication over email. A.M. acknowledges Ministry of Education, India for funding. S.D. would like to acknowledge the MATRICS
grant (MTR/ 2019/001 043) from the Science and Engineering Research Board (SERB) for funding.

\bibliography{Paper.bib}

\onecolumngrid

\onecolumngrid

\appendix
\renewcommand\appendixname{Appendix}


\section{Matrix formalism}
\label{matrixFormalism}
To have a clear physical insight into different semi-classical paths that give rise to degeneracy-lifted ABS and the thermoelectric response of a JJ, we shall be using the matrix method as discussed by A. Kundu et. al. \cite{kundu_PRB_82_155441}.

Let $\Psi_{qp[N]}^{e+(-)}$ and $\Psi_{qp[N]}^{h+(-)}$ denote forward (backward) moving electron-like QP and forward (backward) moving hole-like QP respectively within the superconducting lead $S_i$ having superconducting phase $\phi_i$ $(i \in \{1,2\})$ [within the normal region]. These wave functions  can be explicitly calculated using the BdG Hamiltonian (\ref{DiracHamiltonian}) in the main text or the BdG Hamiltonian with quadratic dispersion and with s-wave or p-wave superconductivity
\begin{align}
	\mathcal{H}_{\eta}= \left( -\dfrac{\hbar^2}{2m}\dfrac{\partial^2}{\partial x^2}-\mu  \right) \tau_z + \Delta^{\eta}(x) (\cos \phi_r \tau_x - \sin \phi_r \tau_y);	\label{HamiltonianSP}
\end{align}
where $\Delta^{\eta}(x)=\Delta_0 [ \Theta(-x-L/2)+\Theta(x-L/2) ]\mathit{f}(\eta)$, $\mathit{f}(\eta)=(-i\partial_x/k_F)^{(1-\eta)/2}$, $\eta=\pm 1$ for s-wave and p-wave superconductivity respectively and $p_F=\hbar k_F=\sqrt{2m \mu}$ is the Fermi momentum.

We first consider two reflection matrices $\mathbb{R}^{\gamma}$, $\gamma \in \{1,2\}$, which describe both Andreev and normal reflections at the normal-superconducting junctions.
\begin{equation*}
	\begin{aligned}[c]
		\mathbb{R}^1 \Psi^{e+}_N &= r^1_{Ahe} \Psi^{h-}_N + r^1_{Nee} \Psi^{e-}_N,\\
		\mathbb{R}^1 \Psi^{h-}_N &= r^1_{Aeh} \Psi^{e+}_N + r^1_{Nhh} \Psi^{h+}_N,
	\end{aligned}
	\qquad
	\begin{aligned}[c]
		\mathbb{R}^2 \Psi^{e-}_N &= r^2_{Ahe} \Psi^{h+}_N + r^2_{Nee} \Psi^{e+}_N,\\
		\mathbb{R}^2 \Psi^{h+}_N &= r^2_{Aeh} \Psi^{e-}_N + r^2_{Nhh} \Psi^{h-}_N,
	\end{aligned}
	\qquad
	\begin{aligned}[c]
		\mathbb{R}^1 \Psi^{e-}_N &= \mathbb{R}^1 \Psi^{h+}_N = \mathbb{R}^2 \Psi^{e+}_N = \mathbb{R}^2 \Psi^{h-}_N = 0,
	\end{aligned}
\end{equation*}
where $r^{\gamma}_{Aqq'}$ and $r^{\gamma}_{Nqq'}$ respectively describe the amplitudes of different Andreev reflections and normal reflections.

To consider the propagation of the wave functions through a length $l$ within the normal region, we consider two propagation matrices, $\mathbb{T}^{\gamma}$ $(\gamma \in \{1,2\})$, such that
\begin{equation*}
	\begin{aligned}[c]
		&\mathbb{T}^1(l) \Psi^{e+}_N|_{x} = \Psi^{e+}_N|_{x+l},\\
		&\mathbb{T}^1(l) \Psi^{h-}_N|_{x} = \Psi^{h-}_N|_{x-l},
	\end{aligned}
	\qquad
	\begin{aligned}[c]
		&\mathbb{T}^2(l) \Psi^{e-}_N|_{x} = \Psi^{e-}_N|_{x-l},\\
		&\mathbb{T}^2(l) \Psi^{h+}_N|_{x} = \Psi^{h+}_N|_{x+l},
	\end{aligned}
	\qquad
	\begin{aligned}[c]
			&\mathbb{T}^1(l) \Psi^{e-}_N = \mathbb{T}^1(l) \Psi^{h+}_N = \mathbb{T}^2(l) \Psi^{e+}_N = \mathbb{T}^2(l) \Psi^{h-}_N =0.
	\end{aligned}
\end{equation*}

For energies above the superconducting gap, two tunneling matrices at the two boundaries, $\mathbb{T}^{L,R}_B$, are defined as
\begin{equation*}
	\begin{aligned}[c]
		&\mathbb{T}_B^L \Psi_{qp}^{e+}[\phi_1] = t_e \Psi^{e+}_N + t_{Ae} \Psi^{h+}_N,\\
		&\mathbb{T}_B^L \Psi_{qp}^{h+}[\phi_1] = t_h \Psi^{h+}_N + t_{Ah} \Psi^{e+}_N,
	\end{aligned}
	\qquad
	\begin{aligned}[c]
		&\mathbb{T}_B^R \Psi^{e+}_N = t_e^{qp} \Psi_{qp}^{e+}[\phi_2] + t_{Ae}^{qp}\Psi_{qp}^{h+}[\phi_2],\\
		&\mathbb{T}_B^R \Psi^{h+}_N = t_h^{qp} \Psi_{qp}^{h+}[\phi_2] + t_{Ah}^{qp} \Psi_{qp}^{e+}[\phi_2],
	\end{aligned}
	\qquad
	\begin{aligned}[c]
		&\mathbb{T}_B^L \Psi_{qp}^{e-} = \mathbb{T}_B^L \Psi_{qp}^{h-} \\
		&= \mathbb{T}_B^R \Psi^{e-}_N= \mathbb{T}_B^R \Psi^{h-}_N = 0.
	\end{aligned}
\end{equation*}

We also consider scattering matrices within the normal region to account for the disorders,
\begin{equation*}
	\begin{aligned}[c]
		\mathscr{T}^e
		\begin{bmatrix}
			\Psi^{e+}|_{-x}	\\
			\Psi^{e-}|_{+x}
		\end{bmatrix}
		&=
		\begin{bmatrix}
			\Psi^{e+}|_{+x}	\\
			\Psi^{e-}|_{-x}
		\end{bmatrix}
	\end{aligned}
	\qquad
	\begin{aligned}[c]
		\mathscr{T}^h
		\begin{bmatrix}
			\Psi^{h-}|_{+x}	\\
			\Psi^{h+}|_{-x}
		\end{bmatrix}
		&=
		\begin{bmatrix}
			\Psi^{h-}|_{-x}	\\
			\Psi^{h+}|_{+x}
		\end{bmatrix}
	\end{aligned}
\end{equation*}
Note that, the matrices $\mathscr{T}^e$ and $\mathscr{T}^h$ are related by the particle-hole symmetry of the corresponding BdG Hamiltonian.

Explicit expressions of the reflection matrices $\mathbb{R}^{\gamma}$ and tunneling matrices $\mathbb{T}^{L,R}_B$ can be obtained by demanding the continuity of the wave functions across the boundaries in case of JJ based on HES or by using the following boundary conditions in case of JJ with quadratic dispersion\cite{tinyukova2019andreev}
\begin{align}
	\dfrac{\hbar^2}{2m} \tau_z \left[ \partial_x^{(\beta)} \Psi_S^{\pm} - \partial_x^{(\beta)} \Psi_N^{\pm} \right] +i \beta \left(\dfrac{1-\eta}{2}\right) \dfrac{\Delta_0}{k_F} \left[ \cos \phi_{\pm} \tau_x - \sin \phi_{\pm} \tau_y \right] \Psi_S^{\pm}=0
\end{align}
where $\beta \in \{0,1\}$; $\eta=1$ for s-wave and $\eta=-1$ for p-wave superconductivity; $\phi_{+}=\phi_2$ and $\phi_{-}=\phi_1$; $\Psi_S$ and $\Psi_N$ are the wave functions in the superconducting and normal regions respectively.

\section{Clean junction}
\label{Appendix_clean_junction}
Andreev bound states are the result of multiple Andreev reflections. There are two ways in which Andreev bound state can be formed as discussed in the main text. We shall describe the same processes here with the help of matrix formalism discussed in \ref{matrixFormalism}.

\textit{(i) Tunneling of a Cooper pair from left to right:} An electron-like quasiparticle starts at $x=-L/2$ (i.e. $\Psi^{e+}_N|_{x=-L/2}$) and propagates through the normal region and reaches at $x=L/2$ (i.e. $\Psi^{e+}_N|_{x=L/2}=\mathbb{T}^{1} \Psi^{e+}_N|_{x=-L/2}$). It Andreev reflects back as a hole with uni-modular amplitude $r^1_{Ahe}$ (i.e. $r^1_{Ahe}\Psi^{h-}_N= \mathbb{R}_A^{1} \Psi^{e+}_N$) by creating a Cooper pair in the superconducting lead 2 (S2). The reflected hole then travels through the normal region and reaches at $x=-L/2$ (i.e. $\Psi^{h-}_N|_{x=-L/2} = \mathbb{T}^{1} \Psi^{h-}_N|_{x=L/2}$). It then again Andreev reflects as an electron with uni-modular amplitude $r^1_{Aeh}$ (i.e. $r^1_{Aeh} \Psi^{e+}_N = \mathbb{R}_A^{1} \Psi^{h-}_N$) by annihilating a Cooper pair in the superconducting lead 1 (S1). Now for $\omega \leq \Delta_0$, matrices $\mathbb{R}^{\gamma}$ and $\mathbb{T}^{\gamma}$ are unitary, so it must be
\begin{align}
	\Psi^{e+}_N|_{x=-L/2} = (\mathbb{R}^{1}\mathbb{T}^{1}\mathbb{R}^{1}\mathbb{T}^{1}) \Psi^{e+}_N|_{x=-L/2}.	\label{conditionABSfirstType}
\end{align}
The corresponding Andreev bound state energy can be obtained by solving the determinant condition
\begin{align}
	\text{det.}(\mathbb{I}_{4\times 4}-\mathbb{R}^{1}\mathbb{T}^{1}\mathbb{R}^{1}\mathbb{T}^{1}) =0,
\end{align}
which gives the ABS energy $\omega_0^{21}$.

\textit{(ii) Tunneling of a Cooper pair from right to left:} If a right-moving hole-like quasiparticle starts from $x=-L/2$ (i.e. $\Psi^{h+}_N|_{x=-L/2}$) and completes the cycle after two Andreev reflections, it can transfer a Cooper pair from S2 to S1
\begin{align}
	\Psi^{h+}_N|_{x=-L/2} = (\mathbb{R}^2\mathbb{T}^2\mathbb{R}^2\mathbb{T}^2) \Psi^{h+}_N|_{x=-L/2}.	\label{conditionABSsecondType}
\end{align}
The corresponding Andreev bound state energy can be obtained by solving the equation
\begin{align}
	\text{det.}(\mathbb{I}_{4\times 4}-\mathbb{R}^{2}\mathbb{T}^{2}\mathbb{R}^{2}\mathbb{T}^{2}) =0,
\end{align}
which gives the ABS energy $\omega_0^{12}$.

Now, tunneling of a quasiparticle with energy $\omega>\Delta_0$ from S1 to S2 can be understood in terms of the matrices $\mathbb{R}^{\gamma}$, $\mathbb{T}^{\gamma}$ and $\mathbb{T}_B^{(L,R)}$.

\textit{(i) Tunneling of an electron (hole)-like quasiparticle from left (right) to right (left):} For a clean junction, an incident electron-like quasiparticle in S1 (i.e. $\Psi_{qp}^{e+}[\phi_1]$) can tunnel into S2 as a electron-like quasiparticle (i.e. $\Psi_{qp}^{e+}[\phi_2]$) either directly or by any even number of Andreev reflections. Mathematically,
\begin{align}
	\chi_{ee}^{21} \Psi_{qp}^{e+}[\phi_2]&=\mathbb{T}_B^R (\mathbb{T}^1 + \mathbb{T}^1 \mathbb{R}^1 \mathbb{T}^1 \mathbb{R}^1 \mathbb{T}^1 + ...)\mathbb{T}_B^L \Psi_{qp}^{e+}[\phi_1] = \mathbb{T}_B^R \mathbb{T}^1(\mathbb{I}-\mathbb{R}^1 \mathbb{T}^1 \mathbb{R}^1 \mathbb{T}^1)^{-1} \mathbb{T}_B^L \Psi_{qp}^{e+}[\phi_1].	\label{electronTransmission}
\end{align}
It is clear from Eq. (\ref{electronTransmission}) and (\ref{conditionABSfirstType}) that the tunneling of an electron-like quasiparticle from S1 to S2 is in correspondence with the Andreev bound state having energy $\omega_0^{21}$. Solving Eq. (\ref{electronTransmission}) we can calculate $\chi_{ee}^{21}$ and hence $\mathcal{T}_{ee}^{21}$.

\textit{(ii) Tunneling of an hole (electron)-like quasiparticle from left (right) to right (left):} Similarly, tunneling of a hole-like quasiparticle from S1 to S2 can be mathematically expressed as
\begin{align}
	\chi_{hh}^{21} \Psi_{qp}^{h+}[\phi_2]&=\mathbb{T}_B^R (\mathbb{T}^2 + \mathbb{T}^2 \mathbb{R}^2 \mathbb{T}^2 \mathbb{R}^2 \mathbb{T}^2 + ...)\mathbb{T}_B^L \Psi_{qp}^{h+}[\phi_1]= \mathbb{T}_B^R \mathbb{T}^2(\mathbb{I}-\mathbb{R}^2 \mathbb{T}^2 \mathbb{R}^2 \mathbb{T}^2)^{-1} \mathbb{T}_B^L \Psi_{qp}^{h+}[\phi_1].	\label{holeTransmission}
\end{align}
A comparison between Eq. (\ref{holeTransmission}) and (\ref{conditionABSsecondType}) clearly indicates the fact that the tunneling of a hole-like quasiparticle from S1 to S2 is in correspondence with the Andreev bound state having energy $\omega_0^{12}$. Solving Eq. (\ref{holeTransmission}) we can calculate $\chi_{hh}^{21}$ and hence $\mathcal{T}_{hh}^{21}$.

\section{Significance of the quantity $(k_e-k_h)L/2$}
\label{ApproximationkL}
We have assumed the doping of the junction is sufficiently high, so let us retain the expressions of $k_e$ and $k_h$ up to the first order of $\omega/\mu$ for quadratic dispersion relation,
\begin{align}
	k_e =\dfrac{\sqrt{2m}}{\hbar}\sqrt{\mu+\omega} \approx \dfrac{\sqrt{2m \mu}}{\hbar} \left( 1+\dfrac{\omega}{2\mu} \right)	\text{	;	}	k_h =\dfrac{\sqrt{2m}}{\hbar}\sqrt{\mu-\omega} \approx \dfrac{\sqrt{2m \mu}}{\hbar} \left( 1-\dfrac{\omega}{2\mu} \right)
\end{align}
Now, we shall consider the length of the junction $L$ to be finite compare to the superconducting coherence length $\xi=\hbar \sqrt{2\mu/m}/\Delta_0$ so let $L=\mathrm{x}\xi$. Now,
\begin{align}
	\dfrac{k_e-k_h}{2}L
	\approx  \dfrac{1}{2} \dfrac{\sqrt{2m \mu}}{\hbar} \left[ \left( 1+\dfrac{\omega}{2\mu} \right)-\left( 1-\dfrac{\omega}{2\mu} \right) \right] \mathrm{x}\xi
	\approx  \dfrac{1}{2} \dfrac{\sqrt{2m \mu}}{\hbar} \dfrac{\omega}{\mu} \left( \mathrm{x} \dfrac{\hbar}{\Delta_0} \sqrt{\dfrac{2\mu}{m}} \right)	
	\approx  \mathrm{x} \dfrac{\omega}{\Delta_0}.
\end{align}
Thus, even for large enough doping, the quantity $(k_e-k_h)L/2$ is of the order of $\omega/\Delta_0$, and thus cannot be neglected.

For linear dispersion relation, $k_e =\frac{\mu+\omega}{\hbar v_F}$ and $k_h =\frac{\mu-\omega}{\hbar v_F}$, hence, here also $\frac{k_e-k_h}{2}L \approx  \mathrm{x} \frac{\omega}{\Delta_0}$.

\section{Presence of a scatterer in the middle of the junction}
\label{Appendix_middle_scatterer}
Starting with an initial state $\left( (\Psi^{e+}_N|_{x=-L/2}), (\Psi^{e-}_N|_{x=L/2}) \right)^T$, it will come back to the same state after a electron scattering followed by an Andreev reflection, a hole scattering and another Andreev reflection. For $\omega<\Delta_0$, these matrices all being unitary, it must be
\begin{align}
	\begin{bmatrix}
		\Psi^{e+}_N|_{x=-L/2}	\\
		\Psi^{e-}_N|_{x=L/2}
	\end{bmatrix}
	=
	\mathbb{R}^A_P \mathscr{T}^h_P \mathbb{R}^A_P \mathscr{T}^e_P
	\begin{bmatrix}
		\Psi^{e+}_N|_{x=-L/2}	\\
		\Psi^{e-}_N|_{x=L/2}
	\end{bmatrix}
\end{align}
where we have defined $\mathbb{M}_P=\mathbb{M} \mathbb{T}^P$. Note that, in the absence of barrier i.e. at $\mathscr{T}^e=\mathscr{T}^h=\mathbb{I}$, all the matrices $\mathbb{T}^P$, $\mathbb{R}^A$, $\mathscr{T}^e$ and $\mathscr{T}^h$ are block diagonal and the aforesaid two types of ABS ($\omega_0^{21}$ and $\omega_0^{12}$) do not interfere. In presence of barrier, finite backscattering (off-diagonal blocks of $\mathscr{T}^e$ and $\mathscr{T}^h$) gives rise to the interference between the two types of ABS ($\omega_0^{21}$ and $\omega_0^{12}$).

ABS energies, in presence of barrier can be obtained by solving the equation
\begin{align}
	\text{det}.\left( \mathbb{I}_{4\times 4}-\mathbb{R}^A_P \mathscr{T}^h_P \mathbb{R}^A_P \mathscr{T}^e_P \right) =0
	\label{EQ63}
\end{align}
Note that, if we had started with the initial state $\left( (\Psi^{h-}|_{x=L/2}), (\Psi^{h+}|_{x=-L/2}) \right)^T$ then Eq. (\ref{EQ63}) would have looked like
\begin{align}
	\text{det}.\left( \mathbb{I}_{4\times 4}-\mathbb{R}^A_P \mathscr{T}^e_P \mathbb{R}^A_P \mathscr{T}^h_P \right) =0	\label{EQ64}
\end{align}
It turns out, the ABS energies, as obtained from (\ref{EQ63}) or (\ref{EQ64}) are same.

For energies $\omega>\Delta_0$, we define the following matrices
\begin{equation*}
	\begin{aligned}[c]
		\mathbb{T}^L =
		\begin{bmatrix}
			\mathbb{T}_B^L	&0	\\
			0	&\mathbb{T}_B^L
		\end{bmatrix}
	\end{aligned}
	\qquad
	\begin{aligned}[c]
		\mathbb{T}^R_e =
		\begin{bmatrix}
			\mathbb{T}_B^R	&0	\\
			0	&0
		\end{bmatrix}
	\end{aligned}
	\qquad
	\begin{aligned}[c]
		\mathbb{T}^R_h =
		\begin{bmatrix}
			0	&0	\\
			0	&\mathbb{T}_B^R
		\end{bmatrix}
	\end{aligned}
\end{equation*}
With this, tunneling of a QP from S1 to S2 can be understood as follows:

\textbf{(i)} An incident electron-like QP in S1 $\left( (\Psi_{qp}^{e+}[\phi_1]),(0) \right)^T$ can tunnel into S2 as an electron-like QP $\left( (\Psi_{qp}^{e+}[\phi_2]),(0) \right)^T$ either directly or by any even number of Andreev reflections whereas tunneling of an electron-like QP from S1 into S2 as an hole like QP $\left( (0),(\Psi_{qp}^{h+}[\phi_2]) \right)^T$ must be mediated by an odd number of Andreev reflections.
\begin{equation*}
	\begin{aligned}[c]
		\chi_{ee}^{21} 
		\begin{bmatrix}
			\Psi_{qp}^{e+}[\phi_2] 	\\
			0
		\end{bmatrix}
		&=\mathbb{T}_e^R \mathbb{T}^P \mathscr{T}^e_P (\mathbb{B}^e)^{-1} \mathbb{T}^L
		\begin{bmatrix}
			\Psi_{qp}^{e+}[\phi_1] 	\\
			0
		\end{bmatrix}
	\end{aligned}
	\qquad
	\begin{aligned}[c]
		\chi_{he}^{21}
		\begin{bmatrix}
			0	\\
			\Psi_{qp}^{h+}[\phi_2]
		\end{bmatrix}
		&= \mathbb{T}_h^R \mathbb{T}^P \mathscr{T}^h_P \mathbb{R}^A_P \mathscr{T}^e_P (\mathbb{B}^e)^{-1}\mathbb{T}^L
		\begin{bmatrix}
			\Psi_{qp}^{e+}[\phi_1]	\\
			0
		\end{bmatrix}
	\end{aligned}
\end{equation*}
where $\mathbb{B}^e=\mathbb{I}_{4\times 4}-\mathbb{R}^A_P \mathscr{T}_P^{h}\mathbb{R}^A_P\mathscr{T}_P^{e}$. Solving above equations we can calculate $\chi_{ee}^{21}$ and $\chi_{he}^{21}$ and hence we $\mathcal{T}_{ee}^{21}$ and $\mathcal{T}_{he}^{21}$.

\textbf{(ii)} Similarly, tunneling of a hole-like QP from S1 $\left( (0),(\Psi_{qp}^{h+}[\phi_1]) \right)^T$ into  S2 as an hole-like QP $\left( (0),(\Psi_{qp}^{h+}[\phi_2]) \right)^T$ can be mediated directly or by any even number of Andreev reflections whereas tunneling of a hole-like QP from S1 into S2 as an electron-like QP $\left( (\Psi_{qp}^{e+}[\phi_2]),(0) \right)^T$ must be mediated by an odd number of Andreev reflections.
\begin{equation*}
	\begin{aligned}[c]
		\chi_{hh}^{21} 
		\begin{bmatrix}
			0	\\
			\Psi_{qp}^{h+}[\phi_2]
		\end{bmatrix}
		&=\mathbb{T}_h^R \mathbb{T}^P \mathscr{T}^h_P (\mathbb{B}^h)^{-1} \mathbb{T}^L
		\begin{bmatrix}
			0	\\
			\Psi_{qp}^{h+}[\phi_1]
		\end{bmatrix}
	\end{aligned}
	\qquad
	\begin{aligned}[c]
		\chi_{eh}^{21}
		\begin{bmatrix}
			\Psi_{qp}^{e+}[\phi_2]	\\
			0
		\end{bmatrix}
		&= \mathbb{T}_e^R \mathbb{T}^P \mathscr{T}^e_P \mathbb{R}^A_P \mathscr{T}^h_P (\mathbb{B}^h)^{-1}\mathbb{T}^L
		\begin{bmatrix}
			0	\\
			\Psi_{qp}^{h+}[\phi_1]
		\end{bmatrix}
	\end{aligned}
\end{equation*}
where $\mathbb{B}^h=\mathbb{I}_{4\times 4}-\mathbb{R}^A_P \mathscr{T}_P^{e}\mathbb{R}^A_P\mathscr{T}_P^{h}$. Solving above equations we can calculate $\mathcal{T}_{hh}^{21}$ and $\mathcal{T}_{eh}^{21}$.


\end{document}